\documentstyle[aps]{revtex}
\begin{document}
\draft
\preprint{IFUAP-02-97}
\title{Constraints on the parameters of the Left Right Mirror Model\\
          \vspace{-8ex}
           \hfill{\normalsize hep-ph/9705478 \\[1ex]}
           \hfill{\normalsize IFUAP-02-97\\}
          \vspace{2ex}}
\author{Victoria E. Cer\'on\thanks{Present address: Departamento de F\'{\i}sica, 
         Centro de Investigaci\'on y de Estudios Avanzados del IPN,
         Apartado Postal 14-740, 07000 M\'exico Distrito Federal, M\'exico.},
        Umberto Cotti\thanks{e-mail: ucotti@sirio.ifuap.buap.mx}, 
        J. Lorenzo Diaz-Cruz\thanks{e-mail: ldiaz@sirio.ifuap.buap.mx}
       }
\address{Instituto de F\'{\i}sica, Universidad Aut\'onoma de Puebla, 
         Apartado Postal J-48, 72570 Puebla Puebla, M\'exico.}
\author{Mario Maya}
\address{Facultad de Ciencias F\'{\i}sico Matem\'aticas,
         Universidad Aut\'onoma de Puebla,
         Apartado Postal 1152, 72570 Puebla Puebla, M\'exico.}
\date{May 30, 1997}
\maketitle

\begin{abstract}
 We study some phenomenological constraints on the parameters of a 
{\sl left right model with mirror fermions} (LRMM) that solves the 
strong CP problem.
 In particular, we evaluate the contribution of mirror neutrinos to
the {\sl invisible Z decay width} ($\rm \Gamma_Z^{inv}$), and we
 find that the present experimental value on $\rm \Gamma_Z^{inv}$, 
can be used to place an upper bound on the Z-Z$'$ mixing angle that 
is consistent with limits obtained previously from other low-energy 
observables.
 In this model the charged fermions that correspond to the 
{\sl standard model} (SM) mix with its mirror counterparts. 
 This mixing, simultaneously with the Z-Z$'$ one, leads to modifications 
of the $\Gamma({\rm Z} \rightarrow f \bar{f})$ decay width.
 By comparing with LEP data, we obtain bounds on the standard-mirror 
lepton mixing angles.
 We also find that the bottom quark mixing parameters can be chosen 
to fit the experimental values of $\rm R_b$, and the resulting 
values for the Z-Z$'$ mixing angle do not agree with previous bounds.
 However, this disagreement disappears if one takes the more recent
ALEPH data. 
\end{abstract}

                      \section{Introduction}
 Although the non-conservation of {\sl parity} (P), is well incorporated in 
the SM of electroweak interactions, it has been considered as 
an unpleasant feature of the model, and many attempts have been made to 
restore it; for instance, in Left-Right symmetric models~\cite{book8} 
the problem is solved through the inclusion of $\rm SU(2)_R$ interactions 
that maintain P-invariance at high energy scales.
 Another interesting solution, due to Lee and Yang~\cite{pr104:254},
restores P by the inclusion of additional (mirror) fermions of opposite 
chirality to the SM ones. 

 Low-energy effects of mirror fermions have been studied since 
then, and limits on their masses exist~\cite{prp186:53}. 
 Applications of models with mirror fermions to solve some 
problems in particle physics have been discussed too.
 For instance, it has been recently found that 
mirror neutrinos can help to alleviate the 
problems that appear in this sector, namely the solar and atmospheric 
neutrino deficits, and the Dark Matter problem~\cite{prd52:6595,prd52:6607}. 
 Other interesting effects of mirror neutrinos in astrophysics
and cosmology are discussed in~\cite{hep-ph:9503481}.
 Mirror particles also appear naturally in many extensions 
of the SM, like GUT and string theories~\cite{prp186:53}.
 On the other hand, in Refs.~\cite{prl67:2765,prd41:1286} it is 
proposed a class of mirror models where the strong CP problem is 
solved.
 Because of the attractive features of these models, it seems 
interesting to test further the predictions for the masses and couplings 
of mirror fermions. 

 In this paper, we are interested in testing the validity of a mirror 
model with gauge group $\rm SU(2)_L \otimes SU(2)_R \otimes U(1)$.
 We calculate the contribution of mirror neutrinos to the 
invisible width of the Z boson, the modifications of the leptonic Z decay
and the possibility to explain the $\rm R_b$ 
($\rm = 
 \frac{\Gamma (Z \rightarrow b \bar{b})}{\Gamma (Z \rightarrow hadrons)}$)
value within this model.

 Because the number of light SM neutrinos is consistent with 
3~\cite{prd54:1}, it is usually said that additional massless neutrinos 
are not allowed. 
 However, it is not known if this statement also applies to the case of 
mirror neutrinos.
 In fact, ref.~\cite{prl67:2765} claims that mirror neutrinos do not 
contribute to the invisible Z-width, however, as it shall be 
explained below, because of Z-Z$'$ mixing, mirror neutrinos do couple 
to the Z boson, and thus can contribute to $\rm \Gamma_Z^{inv}$.

 Because of the non-observability at LEP of the decays
Z$ \rightarrow \overline{\!\widehat{f}} \widehat{f}$, the masses of 
charged mirror fermions must be  above $\case{M_{\rm Z}}{2}$,
furthermore, at the Tevatron Collider it should be possible to put 
a higher limit, similar to the top mass, if no new signal is observed. 
 On the other hand the properties of the Z-boson have been measured at 
LEP with a high precision, which has made possible to test the SM at the 
level of radiative corrections~\cite{prp216:253}, and also to constrain 
the presence of new physics. 

 The organization of this paper is as follows: we shall present in detail 
the simplest model that solve the strong CP 
problem~\cite{prl67:2765,prd41:1286} in the next section, giving
particular emphasis to incorporate the mixing effects into a low-energy
effective lagrangian using the formalism of Ref.\cite{prd55:2998}.
 Sec.~\ref{znunu} contains the discussion of the invisible Z decay.
 Sec.~\ref{zee} is devoted to study the bounds on the parameters obtained 
from the decay $\rm Z \rightarrow e^+ e^-$ and in Sec.~\ref{zbb} we 
discuss the decay $\rm Z \rightarrow b \bar{b}$.
 Finally Sec.~\ref{concl} will contain our conclusions.

              \section{The left-right mirror model (LRMM)}
 The strong CP problem is associated to the suppression of the $\theta$ 
term that breaks P and CP symmetries of the QCD lagrangian. 
 The Peccei-Quinn solution to the strong CP problem, predicts a new 
pseudo-goldstone boson, the axion,  which so far has not been 
observed~\cite{prl38:1440}.
 On the other hand Barr et al.~\cite{prl67:2765,prd41:1286}, proposed a 
class of model that offer a solution to the strong CP problem based on the 
complete invariance of the theory under P.
 In the simplest L-R model of this class, the electroweak group is extended 
to  $\rm SU(2)_L \otimes SU(2)_R \otimes U(1)$, and the matter content of 
the theory is also enlarged by including new fermion fields with mirror
properties.

 Thus, in the LRMM that will be studied here, the right-handed 
(left-handed) components of mirror fermions will transform as doublets 
(singlets) under $\rm SU(2)_R$. 
 The SM fermions are singlets under $\rm SU(2)_R$, whereas the 
right-handed mirrors are also singlets under $\rm SU(2)_L$. 
 Mirror and SM fermions will share hypercharge and color interactions%
\footnote{We shall use the term ``SM fields'' to denote the fermions
with the same chiralities as in the SM, eventhough they will be the same only in some approximation.}.

 The first family of the leptonic sector will be written as follows:
\begin{equation}
 \rm {\bbox l}_{e L}^o = 
 \left(
  \begin{array}{c}
   \rm \nu^o_{\rm e} \\
   \rm   e^o
  \end{array}
 \right)_{\rm L} 
 \equiv \rm {\bbox l}^o_{\rm e}, 
 \hspace{4ex}
 \rm e^o_R  \equiv e^o,
 \hspace{8ex}
 \rm \widehat{\bbox l}_{e R}^o = 
 \left( 
  \begin{array}{c}
   \rm \widehat{\nu}^o_{\rm e} \\ 
   \rm \widehat{e}^o
  \end{array}
 \right)_{\rm R} 
 \equiv \rm \widehat{\bbox l}^o_{\rm e}, \hspace{4ex}
 \rm \widehat{e}^o_L \equiv \widehat{e}^o.
\label{lep}
\end{equation}
The superscript ($ \rm {}^o$) denotes weak eigenstates, and the hat symbol 
( $\widehat{}$ ) is associated to mirror particles.
 Because the model does not contain left-handed mirror neutrinos,
they have to be massless. 
 Similar terms can be written for the quarks and the other families.

                   \subsection{Symmetry breaking}
 In order to realize the breaking of the gauge symmetry, two Higgs doublets 
are included, the SM one ($\bbox{\phi}$) and its mirror partner 
($\widehat{\bbox{\phi}}$). 
 The potential of the model can be written in such a way that the 
v.e.v.'s of the Higgs fields are:
\begin{equation}
 \bbox{\phi} = \case{1}{\sqrt{2}} 
 \left( 
  \begin{array}{c}
   0 \\ 
   v
  \end{array}
 \right),
 \hspace{8ex}
 \widehat{\bbox{\phi}} = \case{1}{\sqrt{2}} 
 \left( 
  \begin{array}{c}
   0 \\ 
   \widehat{v}
  \end{array}
 \right) .
\label{vac}
\end{equation}

                    \subsubsection{Gauge boson masses}
 The mass matrix for the gauge bosons is obtained from the scalar
 lagrangian, namely:
\begin{equation}
 {\cal L}^{\rm kin} = 
 \left(
  \bbox{D}_\mu \bbox{\phi}
 \right)^{\dag}
 \left(
  \bbox{D}^\mu \bbox{\phi}
 \right) + 
 \left(
  \widehat{\bbox{D}}_{\mu} \widehat{\bbox{\phi}}
 \right)^{\dag}
 \left(
  \widehat{\bbox{D}}^\mu \widehat{\bbox{\phi}}
 \right),
\label{lmasa}
\end{equation}
where $\bbox{D}_\mu$ is the covariant derivative associated to the SM, 
and $\widehat{\bbox{D}}_\mu$ is the one associated with the mirror part. 

 After the substitution of the v.e.v.'s in the lagrangian we obtain the 
expressions for the mass matrices, which could be non-diagonal.
 The vector bosons will be denoted as: $\rm W^{\pm}, Z$, which 
correspond to the SM ones, i.e. those associated to 
$\rm SU(2)_L \otimes U(1)_Y$, whereas the mirror gauge bosons will be 
denoted by $\rm \widehat{W}^\pm, \widehat{Z}$. 
 The mass matrix for the charged gauge bosons is diagonal, with mass values:
$M_{\rm W} = \frac{1}{2}vg$, and 
$M_{\widehat{\rm W}} = \frac{1}{2} \widehat{v} \widehat{g}$, where 
$g$ and $\widehat{g}$ are the coupling constant of the $\rm SU(2)_L$ and 
$\rm SU(2)_R$ gauge group respectively. 
 P invariance requires $g = \widehat{g}$, however in this section we 
shall write the equations with general $g$, $\widehat{g}$, since this
could arise in more general models.

 The mass matrix for the neutral components 
($\rm W_\mu^3, \widehat{W}_\mu^3, B_\mu $)
is not diagonal:
\begin{equation}
 \rm {\cal L}^m_{Z \widehat{Z} A} = \case{1}{8} 
 \left( 
  \rm W_{\mu}^{3}, \widehat{W}_{\mu}^{3},B_{\mu}
 \right) 
 \left( 
  \begin{array}{ccc}
   g^2 v^2         &         0                 & -g \widehat{g} v^2           \\
      0            & g^2 \widehat{v}^2         & -g \widehat{g} \widehat{v}^2 \\
  -g\widehat{g}v^2 &-g\widehat{g}\widehat{v}^2 & g^2(v^2 + \widehat{v}^2)
  \end{array}
 \right) 
 \left(
  \begin{array}{c}
   \rm W^{3}_{\mu}       \\
   \rm \widehat{W}^{3}_{\mu} \\
   \rm B_{\mu}           \\
  \end{array}
 \right).
\end{equation}
 This rank 2 matrix, with non-trivial eigenvalues
\begin{equation}
 M_{\rm Z,\widehat{Z}} = 
 \frac{1}{2}
  \left(
   g^{2} + \widehat{g}^2
  \right)
  \sigma \mp \frac{1}{2} 
  \sqrt{ 
   \left(
    g^{2} + \widehat{g}^2
   \right)^2 \sigma^2 
   - 4g^2 v^2 \widehat{v}^2
   \left(
    g^2 + 2\widehat{g}^2
   \right), 
  }
\end{equation}
and with $\sigma = v^2 + \widehat{v}^2$,
can be diagonalized by an orthogonal transformation {\sf R},
relating the weak 
($ \rm W_{\mu}^{3}, \widehat{W}_{\mu}^{3},B_{\mu} $)
and mass eigenstates basis
($ \rm Z_{\mu}, \widehat{Z}_{\mu},A_{\mu} $)~\cite{npb153:334,npb363:385}
which will be obtained in Sec.~\ref{interacc}.

                \subsubsection{Charged fermion masses}
 The mass lagrangian for the charged fermionic sector includes the 
ordinary Yukawa terms and its mirror partners, however the model allows 
mixing terms between ordinary and mirror fermions singlets, the 
$\rm \lambda_{\hat{e} e}$ terms that are written below for the first family:
\begin{eqnarray}
 \rm {\cal L}^m_{\hat{e} e} & = & 
 \rm \lambda_{ee}  \overline{\bbox{l}^o_e} \bbox{\phi} e^o  +
 \lambda_{\hat{e}\hat{e}} \overline{\widehat{\bbox{l}}^o_e} 
  \widehat{\!\bbox{\phi}} 
 \widehat{e}^o 
 + \lambda_{\hat{e} e} \overline{\widehat{e}^o} e^o + h.c. .
\label{yuka}
\end{eqnarray}
 To diagonalize the mass matrix, one introduces two mixing angles. 
 Thus, for a flavor $f$, one has:
\begin{eqnarray}
 \left( 
  \begin{array}{c}
   f^{\rm o} \\
   \widehat{f}^{\rm o}
  \end{array}
 \right)_{\rm L,R}  & = &
 \left(   
  \begin{array}{cc}
   \cos  \xi^f &  \sin \xi^f \\[1ex]
   -\sin \xi^f &  \cos \xi^f
  \end{array} 
 \right)_{\rm L,R} 
 \left( 
  \begin{array}{c}
   f          \\
   \widehat{f}
  \end{array} 
 \right)_{\rm L, R}
\label{amex}
\end{eqnarray}
 $f_{\rm L,R}$ will be identified as the L- and R-handed components of the 
ordinary SM fermions, whereas $\widehat{f}_{\rm L,R}$ will correspond to the 
new ones. 
 The mixing angles $\xi^f_{\rm L,R}$ are associated to each flavor $f$.

                 \subsection{The interaction lagrangian}
\label{interacc}
 The gauge interactions of quarks and leptons, can be obtained from the 
lagrangian:
\begin{equation}
 \rm {\cal L}^{int} 
 = 
 \overline{\bbox{\psi}} i \gamma^{\mu} \bbox{D}_{\mu} \bbox{\psi}
 +
 \overline{\!\widehat{\bbox{\psi}}} 
 i \gamma^{\mu} \widehat{\bbox{D}}_{\mu} \widehat{\bbox{\psi}} .
\label{l.int}
\end{equation}
 Following Refs.~\cite{prd55:2998,prd38:886}, grouping all fermions of 
a given electric charge $q$ and a given helicity 
$a$ = L, R in a $n_a + m_a$ vector column of $n_a$ ordinary (O) 
and $m_a$ exotic (E) gauge eigenstates
$\bbox{\psi}_a^{\rm o} = (\bbox{\psi}_{\rm O}, \bbox{\psi}_{\rm E})_a^\top$ 
one finds for the neutral currents term 
\begin{eqnarray}
 -\cal{L}^{\rm nc} &=& 
  \sum_{a={\rm L,R}} \bar{\bbox{\psi}}^{\rm o}_{a} 
  \gamma^{\mu} 
 \left(
  g{\sf T}_{3a}, 
  \widehat{g}\widehat{{\sf T}}_{3a}, 
  g'\frac{{\sf Y}_a}{2}
 \right) 
 \bbox{\psi}^{\rm o}_{a}
  \left(
  \begin{array}{l}
   {\rm W}^3           \\
   {\rm \widehat{W}}^3 \\
   \,{\rm B}           \\
  \end{array}
  \right)_{\mu} \label{general1} 
 \nonumber \\
 &=&
  \sum_{a={\rm L,R}} \bar{\bbox{\psi}}_a \gamma^{\mu}
  {\sf U}^{\dag}_a
  \left(
   g{\sf T}_{3a}, 
   \widehat{g}\widehat{{\sf T}}_{3a},
   g' \frac{{\sf Y}_a}{2}
  \right)
  {\sf U}_a
  \bbox{\psi}_a {\sf R}
  \left(
   \begin{array}{l}
    {\rm Z}  \\
    {\rm Z'} \\
    {\rm A}  \\
   \end{array}
  \right)_{\mu} \label{general2}
\end{eqnarray}
where $g'$ is the coupling constant of the U(1) gauge group, 
\begin{equation}
 {\sf R} = 
 {\sf R}_{\rm Z'A}(\beta)
 {\sf R}_{\rm ZA}(\theta_{\rm w}) {\sf R}_{\rm ZZ'}(\alpha) =
 \left( 
  \begin{array}{ccc}
   \rm  c_{\theta_w} c_\alpha & 
   \rm  c_{\theta_w} s_\alpha & 
   \rm  s_{\theta_w}          \\[2ex]
   \rm -s_\alpha c_\beta  - c_\alpha s_{\theta_w} s_\beta &
   \rm  c_\beta  c_\alpha - s_\alpha s_{\theta_w} s_\beta &   
   \rm  s_\beta  c_{\theta_w}                              \\[2ex]
   \rm  s_\alpha s_\beta - c_\alpha s_{\theta_w} c_\beta &
   \rm -c_\alpha s_\beta - s_\alpha s_{\theta_w} c_\beta &
   \rm  c_\beta  c_{\theta_w}                              \\[2ex]   
  \end{array}
 \right),
\label{orthmatr}
\end{equation}
$\rm \theta_w$, $\alpha$ y $\beta$ are the rotation angles between 
the Z-A, Z-Z$'$ and Z$'$-A gauge bosons respectively.
 The ${\sf U}_a$ matrices are the ones which relate the gauge and the 
corresponding mass eigenstates of the fermion fields,
and ${\sf T}_{3a}$, $\widehat{{\sf T}}_{3a}$ and $\widehat{\sf Y}$
are generators of the $\rm SU(2)_L$, $\rm SU(2)_R$ and U(1) 
respectively.

 The orthogonal matrix in Eq.~(\ref{orthmatr}) can be written in terms of only 
two angles by imposing the electromagnetic coupling relation 
\begin{equation}
 g {\rm s}_{\theta_w} {\sf T}_{3} + 
 \widehat{g} {\rm s}_\beta {\rm c}_{\theta_w} \widehat{\sf T}_{3} +
 g' {\rm c}_\beta {\rm c}_{\theta_w} \frac{{\sf Y}}{2}
 \,=\,
 {\sf Q}e
\end{equation}
whose simplest solution is
\begin{equation}
\left\{
\begin{array}{l}
  g {\rm s}_{\theta_w} 
   \,=\, \widehat{g} {\rm s}_\beta {\rm c}_{\theta_w} 
   \,=\, g' {\rm c}_\beta {\rm c}_{\theta_w} 
   \,=\, e  
  \\[2ex]
   {\sf T}_{3} + \widehat{\sf T}_{3} + \frac{{\sf Y}}{2} 
   \,=\, {\sf Q} .
\end{array}
 \right.
\label{simpsol}
\end{equation}
 From Eqs.~(\ref{simpsol}), defining $\lambda \widehat{g} \equiv g$, 
the following relation holds
\begin{equation}
 \rm c_\beta c_{\theta_w}
 = 
 \sqrt{c_{\theta_w}^2 - \lambda^2 s_{\theta_w}^2}
 \equiv
 r_{\theta_w},
\end{equation}
and makes possible to rewrite {\sf R} as a function of two mixing
angles and the coupling constant ratio:
\begin{equation}
 {\sf R} =
 \left( 
  \begin{array}{ccc}
   \rm c_{\theta_w} c_\alpha
  &  
   \rm c_{\theta_w} s_\alpha 
  & 
   \rm  s_{\theta_w}  
  \\[2ex]
   \rm - \frac{1}{c_{\theta_w}} (s_\alpha r_{\theta_w} + \lambda c_\alpha s^2_{\theta_w})
  &  
   \rm \frac{1}{c_{\theta_w}} (  c_\alpha r_{\theta_w} - \lambda s_\alpha s^2_{\theta_w})
  &        
   \rm \lambda s_{\theta_w}  
  \\[2ex]
   \rm  t_{\theta_w} (\lambda s_\alpha - r_{\theta_w} c_\alpha) 
  &  
   \rm -t_{\theta_w} (\lambda c_\alpha + r_{\theta_w} s_\alpha) 
  & 
   \rm r_{\theta_w}
  \end{array}
 \right).
\end{equation}
 Concentrating our attention on the neutral current sector for the 
Z gauge boson we find that the full lagrangian is
\begin{eqnarray}
 -\cal{L}^{\rm nc}_{\rm Z} &=& 
 \sum_{a={\rm L,R}} 
 \bar{\bbox{\psi}}_{a} 
 \gamma^{\mu} 
 {\sf U}^{\dag}_a
 \left(
  g{\sf T}_{3a}, 
  \widehat{g} \widehat{\sf T}_{3a}, 
  g' \frac{{\sf Y}_a}{2}  
 \right) 
 {\sf U}_a
 \bbox{\psi}_{a}
  \left(
  \begin{array}{c}
   \rm c_{\theta_w} c_\alpha \\
   \rm - \frac{1}{c_{\theta_w}} (s_\alpha r_{\theta_w} + \lambda c_\alpha s^2_{\theta_w})\\
   \rm  t_{\theta_w} (\lambda s_\alpha - c_\alpha r_{\theta_w}) \\
  \end{array}
  \right) 
  {\rm Z}_\mu \label{general3} 
 \nonumber \\[2ex]
 &=&
  \frac{e}{\rm s_{\theta_w} c_{\theta_w}}
  \sum_{a={\rm L,R}} 
  \bar{\bbox{\psi}}_a 
  \gamma^{\mu}
  {\sf U}^{\dag}_a
  \left[
   \left(
    \rm c_\alpha - \frac{s^2_{\theta_w}}{r'_{\theta_w}} s_\alpha
   \right)
   {\sf T}_{3a} - 
   \rm \frac{c^2_{\theta_w}}{\lambda^2 r'_{\theta_w}} s_\alpha 
    \widehat{\sf T}_{3a} +
   s^2_{\theta_w}
   \left(
    \frac{s_\alpha}{r'_{\theta_w}} - c_\alpha
   \right)
   {\sf Q}
  \right]
  {\sf U}_a
  \bbox{\psi}_a 
  {\rm Z_\mu} ,
  \label{general4}
\end{eqnarray}
with $\rm \lambda r'_{\theta_w} \equiv r_{\theta_w}$.

                    \section{Phenomenology of the LRMM}
\label{phlrmm}

                      \subsection{The Invisible Z width}
\label{znunu}
 We are now interested in the interactions of 
ordinary and mirror 
neutrinos, for which there are no mixing terms 
(${\sf U}_a = {\sf 1}$), 
the lagrangian for the Z boson and the neutrinos is then:
\begin{eqnarray}
 -{\cal L}^{(0)}_{\rm Z} 
 &=& 
 \frac{e}{\rm s_{\theta_w} c_{\theta_w}}
 \left[
  \frac{1}{2}
  \bar{\bbox{\psi}}_{\rm L}^{(0)} 
  \gamma^{\mu}
  \left(
   \rm c_\alpha - 
   \frac{s^2_{\theta_w}}{r'_{\theta_w}} s_\alpha
  \right)
  \bbox{\psi}_{\rm L}^{(0)}
  -\frac{1}{2}
  \bar{\widehat{\bbox{\psi}}}_{\rm R}^{(0)}
  \gamma^{\mu}   
  \rm \frac{c^2_{\theta_w}}{\lambda^2 r'_{\theta_w}} s_\alpha 
  \widehat{\bbox{\psi}}_{\rm R}^{(0)}
 \right]
 {\rm Z_\mu} ,  
\end{eqnarray}
 where the charge multiplet $\bbox{\psi}_{\rm L}^{(0)}$ and 
$\widehat{\bbox{\psi}}_{\rm R}^{(0)}$ contain the three standard and 
three mirror neutrinos respectively.
 In the limit $\alpha \rightarrow 0$ (no Z-Z$'$ mixing) we recover the SM 
expression which has been used to extract the number of flavors of light 
neutrinos and is interpreted as a limit on the number of 
families~\cite{prd54:1}.
 Then it is said that there is no room for additional massless neutrinos. 
 However, mirror neutrinos also contribute to the invisible
Z decay; the expression for the total invisible width is 
(for $\lambda =1$):
\begin{eqnarray}
 \rm \Gamma^{inv}_Z &=& 
  \Gamma \left(
          \rm Z \rightarrow \bar{\nu} \nu 
         \right) +
  \Gamma \left(
          \rm Z \rightarrow \overline{\widehat{\nu}} \widehat{\nu} 
         \right) 
  \,=\,
  \frac{{\rm G_F} M^3_{\rm Z}}{4 \sqrt{2} \pi}
  \left(
   \left|
    \rm c_\alpha - \frac{s^2_{\theta_w}}{r_{\theta_w}} s_\alpha
   \right|^2 +
   \left|
    \rm  \frac{c^2_{\theta_w}}{r_{\theta_w}} s_\alpha
   \right|^2
  \right) 
  \nonumber \\[2ex]
  &=&
  \frac{{\rm G_F} M^3_{\rm Z}}{4 \sqrt{2} \pi}  
  \left(
   \rm 1 - 2 \frac{s^2_{\theta_w}}{r_{\theta_w}} \alpha 
       + 2 \frac{s^4_{\theta_w}}{r^2_{\theta_w}} \alpha^2 
       + O(\alpha^3)
  \right),
\end{eqnarray}
thus, the contribution of mirror neutrinos to the
$\rm \Gamma^{inv}_Z$, appears up to the $\alpha^2$ term and
could exclude the model if it goes beyond the experimental value.

 The experimental Z width is $498.3 \pm 4.2$ MeV,
whereas the SM value is $497.64 \pm 0.11$ MeV~\cite{prd54:1},
then, in order for the total contribution not to exceed the 
experimental data, the mixing angle has to satisfy: 
\begin{equation}
 - 1.576 \times 10^{-2} < \alpha < 1.087 \times 10^{-2}, 
\end{equation}
within one standard deviation.

 Comparing this limit on $\alpha$ with the ones obtained 
from other observables, one can see that they are in agreement. 
 For instance, the Z-Z$'$ mixing that appears in the model will produce a 
deviation from unity on the $\rho$ parameter, as given by:
\begin{equation}
 \Delta \rho = \sin^2 \alpha 
 \left( 
  \frac{M^2_{\widehat{\rm Z}}}{M^2_{\rm Z}} - 1 
 \right),
\end{equation}
as has been discussed in the literature for other models~\cite{smtest}, 
which leads to $\rm s^2_\alpha < 5 \times 10^{-4}$. 
 Thus, our limit for $\alpha$ is consistent with this value,
and the LRMM model survives this test.

  \subsection{The $\protect\bbox{\rm Z \rightarrow e \bar{e}}$ decay}
\label{zee}
 Considering only mixing between leptons with their mirror partners
the width is (for $\lambda =1$)
\begin{eqnarray}
  \Gamma \left(
          \rm Z \rightarrow \bar{e} e 
         \right)
  &=&
  \frac{{\rm G_F} M^3_{\rm Z}}{3 \sqrt{2} \pi}
  \left(
   \left|
    \frac{1}{2}
    \left(
     \rm c_\alpha - \frac{s^2_{\theta_w}}{r_{\theta_w}} s_\alpha
    \right)
    \rm c^2_{\xi_L} +
    \rm s^2_{\theta_w}
    \left(
     \frac{s_\alpha}{r_{\theta_w}} - c_\alpha
    \right)
   \right|^2 +
   \left|
    \rm
    \frac{1}{2}
    \frac{c^2_{\theta_w}}{r_{\theta_w}}
    s_\alpha
    s^2_{\xi_R} -
    \rm s^2_{\theta_w}
    \left(
     \frac{s_\alpha}{r_{\theta_w}} - c_\alpha
    \right)
   \right|^2
  \right)
\end{eqnarray}
where $\xi_a$ ($a$ = L, R) and $\alpha$ are the mixing angle between 
e-$\widehat{\rm e}$, and Z-Z$'$ respectively.
 Up to the third order in the mixing angles the physics beyond the 
standard model is manifest through the $\rm \xi_L$ and $\alpha$
parameters as 
\begin{eqnarray}
  \rm B 
  \left(
   \rm Z \rightarrow \bar{e} e 
  \right)
  &=&
  \frac{1}{\Gamma_{\rm tot}}
  \frac{{\rm G_F} M^3_{\rm Z}}{3 \sqrt{2} \pi}
  \left[
   \left(
    \rm -\frac{1}{2} + s^2_{\theta_w}
   \right)^2 
   +
   \rm s^4_{\theta_w} 
   +
   \frac{s^2_{\theta_w}}{r_{\theta_w}}
   \left(
    \rm \frac{1}{2} - 3s^2_{\theta_w}
   \right)
   \alpha
  \right. 
   \nonumber \\[2ex]
   &&
  \left.
   +
   \left(
    \rm -\frac{1}{4} 
    + 
    s^2_{\theta_w}
    +
    s^4_{\theta_w}
    \left(
     \frac{5}{4 r^2_{\theta_w}} - 2
    \right)
   \right)   
   \alpha^2
   +
   \left(
    \rm -\frac{1}{2} + s^2_{\theta_w}
   \right)
   \rm \xi_L^2 
   +
   O(3)
 \right].
\end{eqnarray}
 This quantity is bounded by the experimental uncertainty in the 
data~\cite{prd54:1} :
\begin{equation}
  \rm B^{exp}
  \left(
   \rm Z \rightarrow e \bar{e}
  \right)
  = 
  (3.366 \pm 0.008) \times 10^{-2}.
\end{equation}

 It is remarkable to see how strong the $\xi_{\rm L}$ and $\alpha$ 
parameters are correlated and as a consequence, we should mention that 
it is not safe to analyze the model taking only one of the limits 
$\alpha \rightarrow 0$ or $\rm \xi_L \rightarrow 0$.

 In Fig.~\ref{znnzee}, we show the allowed region in the 
$\alpha$-$\rm \xi_L$ plane,
which includes also 
the constraints obtained from the previous section:
the darker region is precisely the intersection of the results from the 
$\rm \Gamma_Z^{inv}$ (dashed lines) and the 
$\rm Z \rightarrow e \bar{e}$ analysis (continuous lines).

\subsection{The $\protect\bbox{\rm Z \rightarrow b \bar{b}}$ decay and 
            $\protect\bbox{\rm R_b}$}
\label{zbb}
 Another interesting application of the LRMM model is to the 
$\rm Z \rightarrow b \bar{b}$ decay.
 Until recently, it was thought that the ratio 
$\rm R_B = 
 \frac{\Gamma (Z \rightarrow b \bar{b})}{\Gamma (Z \rightarrow hadrons)}$
was in disagreement with the SM prediction.
 However, recent data~\cite{ppe-96-183} seems to favor again the SM, 
although some small discrepancy survives.
 In an extensive study of Bamert et al.~\cite{prd54:4275} it was 
identified a large class of models where the $\rm R_b$ ``crisis'' 
(old data) was solved.
 The solution appeared on three main classes: 
(i) tree-level modification of the b-vertices,
(ii) tree level top modification,
(iii) general loop effects.
 However when the models are discussed in detail, for particular 
cases, it may happen that those solution do not satisfy the 
requirements arising from other sectors.

 If, model independently, we consider that the only contribution to the 
shift in $\rm R_b$ is given by the shift in the 
$\rm Zb \bar{b}$ couplings, through the width
$\rm \Gamma_b \equiv \Gamma (Z \rightarrow b \bar{b})$~\cite{prl78:2902},
then the general expression relating the new physics 
($\rm R_b$, $\rm \Gamma_b$) with the SM one
($\rm R_b^{SM}$, $\rm \Gamma_b^{SM}$) is:
\begin{equation}
 \rm
 R_b = \frac{\Gamma_b}{\Gamma_b^{SM} 
                       \left( 
                        \displaystyle \frac{1}{R_b^{SM}} - 1
                       \right)
                        + \Gamma_b
                      }.
\end{equation}
 In order to compare with the data, one can write the width 
$\Gamma (\rm Z \rightarrow \bar{b} b)$ in the LRMM as follows:
\begin{eqnarray}
  \Gamma \left(
          \rm Z \rightarrow \bar{b} b
         \right)
  &=&
  \frac{{\rm G_F} M^3_{\rm Z}}{ \sqrt{2} \pi}
  \left(
   \left|
    \frac{1}{2}
    \left(
     \rm c_\alpha - \frac{s^2_{\theta_w}}{r_{\theta_w}} s_\alpha
    \right)
    \rm c^2_{\eta_L} +
    \frac{\rm s^2_{\theta_w}}{3}
    \left(
     \frac{s_\alpha}{r_{\theta_w}} - c_\alpha
    \right)
   \right|^2 +
   \left|
    \rm
    \frac{1}{2}
    \frac{c^2_{\theta_w}}{r_{\theta_w}}
    s_\alpha
    s^2_{\eta_R} -
    \frac{\rm s^2_{\theta_w}}{3}
    \left(
     \frac{s_\alpha}{r_{\theta_w}} - c_\alpha
    \right)
   \right|^2
  \right),
\end{eqnarray}
 where $\eta_a$ ($a$ = L, R) are the mixing angle between 
b-$\widehat{\rm b}$.
 We show in Fig.~\ref{znnzbb}, including terms up to third order in 
the mixing angles, the constraints that arise for our model 
from b-$\rm \widehat{b}$ and Z-Z$'$ mixing.
 The results are such that the mixing parameters appear to shown 
a slightly disagreement with the ones obtained from 
$\rm \Gamma^{inv}$ and the Z-Z$'$ mixing.

 However it should be noted that if the most recent data on 
$\rm R_b$~\cite{ppe-97-017,ppe-97-018} are confirmed, this small
discrepancy disappear and then the $\rm R_b$ ``crisis'' will be 
solved for both the SM and the LRMM.

                            \section{Conclusions}
\label{concl}
 We have studied some phenomenological consequences of a left-right model 
that includes mirror fermions, which gives a solution to the strong 
CP problem. 
 In particular, we have calculated the contribution of mirror neutrinos to 
the invisible Z-width. 
 Mirror neutrinos are massless, first because there is only one chirality 
and it is not possible to write a Dirac mass, and second because 
the Higgs sector does not include a Majorana mass term.
 It is found that the current experimental value for the invisible width 
of the Z boson and the leptonic decays, implies a bound on the Z-Z$'$ 
mixing angle, that is however consistent with the limits obtained from 
other low-energy observables.
 Similar conclusions can be reached for the width 
$\rm \Gamma (Z \rightarrow b \widehat{b})$ and 
$\Gamma ({\rm Z} \rightarrow l^+ l^-)$, although the data show
some deficit on the Z-Z$'$ mixing angle. 
 However more recent results imply that the differences tend to 
disappear.
 Thus, this type of model is consistent with experiment at present.
 However, further improvements on the precision of the data can be 
used to search for a first evidence of the model, or to discard it.

                        \acknowledgments
We acknowledge financial support from CONACyT and SNI in Mexico.


\begingroup\raggedright\endgroup


\begin{figure}
 \caption{Allowed ranges of parameters in the $\alpha$-$\rm \xi_L$ plane.
           The region between the dashed lines is obtained from the 
          $\rm \Gamma_Z^{inv}$ and the region between the continous 
          lines from the $\rm Z \rightarrow e \bar{e}$ data. 
           The darker region represent the allowed intersection which 
          simultaneously consider the two processes.
          \label{znnzee}
          }
\end{figure}

\begin{figure}
 \caption{Allowed regions in the $\alpha$-$\rm \eta_L$ plane 
          obtained from $\rm \Gamma_Z^{inv}$ (dashed lines) 
          and $\rm Z \rightarrow b\bar{b}$ (continous lines) data.
           Unlike the situation of Fig.~\protect\ref{znnzee}, there 
          is no region in the plane where there is simultaneous 
          agreeement with data from both processes.
          \label{znnzbb}
          }
\end{figure}

\begin{thebibliography}{10}
\def\href#1#2{#2} 
\bibitem{book8}
R.~N. Mohapatra, {\em Unification and Supersymmetry}.
\newblock Graduate Texts in Contemporary Physics. Springer Verlag, New York,
  NY, USA, second~ed., 1992.

\bibitem{pr104:254}
T.~D. Lee and C.~N. Yang, ``Question of parity conservation in weak
  interactions,'' {\em Phys.Rev.} {\bf 104} (1956) 254.

\bibitem{prp186:53}
J.~Maalampi and M.~Roos, ``Physics of mirror fermions,'' {\em Phys.Rept.} {\bf
  186} (1990) 53.

\bibitem{prd52:6595}
R.~Foot and R.~R. Volkas, ``{Neutrino physics and the mirror world: How exact
  parity symmetry explains the solar neutrino deficit, the atmospheric neutrino
  anomaly and the LSND experiment},'' {\em Phys.Rev.} {\bf D52} (1995)
  6595--6606, \href{http://xxx.lanl.gov/abs/hep-ph/9505359}{{\tt
  hep-ph/9505359}}.

\bibitem{prd52:6607}
Z.~G. Berezhiani and R.~N. Mohapatra, ``Reconciling present neutrino puzzles:
  Sterile neutrinos as mirror neutrinos,'' {\em Phys.Rev.} {\bf D52} (1995)
  6607--6611, \href{http://xxx.lanl.gov/abs/hep-ph/9505385}{{\tt
  hep-ph/9505385}}.

\bibitem{hep-ph:9503481}
Z.~K. Silagadze, ``Neutrino mass and the mirror universe,''
  \href{http://xxx.lanl.gov/abs/hep-ph/9503481}{{\tt hep-ph/9503481}}.

\bibitem{prl67:2765}
S.~M. Barr, D.~Chang, and G.~Senjanovi\'c, ``{Strong CP problem and parity},''
  {\em Phys.Rev.Lett.} {\bf 67} (1991) 2765.

\bibitem{prd41:1286}
K.~S. Babu and R.~N. Mohapatra, ``{A solution to the strong CP problem without
  an axion},'' {\em Phys.Rev.} {\bf D41} (1990) 1286.

\bibitem{prd54:1}
{\bf {Particle Data Group}} Collaboration, R.~M. Barnett {\em et.~al.},
  ``Review of particle physics,'' {\em Phys. Rev.} {\bf D54} (1996) 1--720.

\bibitem{prp216:253}
{\bf ALEPH} Collaboration, {D. Decamp, and others}, ``{Searches for new
  particles in Z decays using the ALEPH detector},'' {\em Phys.Rept.} {\bf 216}
  (1992) 253--340.

\bibitem{prd55:2998}
U.~Cotti and A.~Zepeda, ``Model-independent analysis of the simultaneous mixing
  of gauge bosons and mixing of fermions,'' {\em Phys. Rev.} {\bf D55} (1997)
  2998--3005, \href{http://xxx.lanl.gov/abs/hep-ph/9611442}{{\tt
  hep-ph/9611442}}.

\bibitem{prl38:1440}
R.~D. Peccei and H.~R. Quinn, ``{CP} conservation in the presence of
  instantons,'' {\em Phys.Rev.Lett.} {\bf 38} (1977) 1440--1443.

\bibitem{npb153:334}
G.~Senjanovi\'c, ``Spontaneous breakdown of parity in a class of gauge
  theories,'' {\em Nucl.Phys.} {\bf B153} (1979) 334.

\bibitem{npb363:385}
J.~Polak and M.~Zralek, ``Updated constraints on the neutral current parameters
  in left-right symmetric model,'' {\em Nucl.Phys.} {\bf B363} (1991) 385.

\bibitem{prd38:886}
P.~Langacker and D.~London, ``Mixing between ordinary and exotic fermions,''
  {\em Phys. Rev.} {\bf D38} (1988) 886--906.

\bibitem{smtest}
W.~Hollik, ``Renormalization of the standard model,'' in {\em Precision tests
  of the standard electroweak model} (P.~Langacker, ed.), vol.~14 of {\em
  Directions in High Energy Physics}, (Singapore), World Scientific, 1995.

\bibitem{ppe-96-183}
{\bf {LEP Electroweak Working Group and the SLD Heavy Flavor Group}}
  Collaboration, D.~Abbaneo {\em et.~al.}, ``A combination of preliminary
  electroweak measurements and constraints on the standard model,''
  CERN-PPE-96-183.

\bibitem{prd54:4275}
P.~Bamert, C.~Burgess, J.~Cline, D.~London, and E.~Nardi, ``{$R_b$ and new
  physics: A comprehensive analysis},'' {\em Phys. Rev.} {\bf D54} (1996)
  4275--4300, \href{http://xxx.lanl.gov/abs/hep-ph/9602438}{{\tt
  hep-ph/9602438}}.

\bibitem{prl78:2902}
J.~Berbab\'eu, D.~Comelli, A.~Pich, and A.~Santamaria, ``Hard $m_t$ corrections
  as a probe of the symmetry breaking sector,'' {\em Phys. Rev. Lett.} {\bf 78}
  (1997) 2902--2905, \href{http://xxx.lanl.gov/abs/hep-ph/9612207}{{\tt
  hep-ph/9612207}}.

\bibitem{ppe-97-017}
{\bf {ALEPH}} Collaboration, R.~Barate {\em et.~al.}, ``{A Measurement of $R_b$
  using a lifetime mass tag},'' CERN-PPE-97-017.

\bibitem{ppe-97-018}
{\bf {ALEPH}} Collaboration, R.~Barate {\em et.~al.}, ``{A Measurement of $R_b$
  using mutually exclusive tags},'' CERN-PPE-97-018.

\end{thebibliography}
\end{document}